# Investigation of the Newly Proposed Carrier-Envelope-Phase Stable Attosecond Pulse Source

Z. Tibai[1], Gy. Tóth[2], Zs. Nagy-Csiha[1], J. A. Fülöp[2,3], G. Almási[1,2] and J. Hebling[1,2,3]
[1]*Institute of Physics, University of Pécs, 7624 Pécs, Hungary*
[2]*MTA-PTE High-Field Terahertz Research Group, 7624 Pécs, Hungary*
[3]*Szentágothai Research Centre, 7624 Pécs, Hungary*

**Abstract:** Practical aspects of the robust method we recently proposed for producing few-cycle attosecond pulses with arbitrary waveform in the extreme ultraviolet spectral range are studied numerically. It is based on the undulator radiation of relativistic ultrathin electron layers produced by inverse free-electron laser process. Optimal conditions for nanobunching are given; attosecond pulse energy and waveform, and their stability are studied. For $K$=0.8 undulator parameter, carrier-envelope-phase stable pulses with >45 nJ energy and 80 as duration at 20 nm, and >250 nJ energy and 240 as duration at 60 nm are predicted with 31 mrad and 13 mrad phase stability, respectively.

## I. INTRODUCTION

In recent years a few phenomena sensitive to the carrier-envelope-phase (CEP) of ultrashort laser pulses were recognized [1,2]. Waveform-controlled few-cycle laser pulses enabled the generation of isolated attosecond pulses in the extreme ultraviolet (EUV) spectral range and their application to the study of electron dynamics in atoms, molecules, and solids [3]. The observation of inner-atomic (strong-field) phenomena and EUV pump—EUV probe measurements require intense CEP-controlled attosecond pulses [4-6]. EUV pump—EUV probe experiments can be carried out at free-electron lasers (FELs) [7,8]; however, the temporal resolution is limited to the few fs regime.

Various schemes, such as the longitudinal space charge amplifier [9], or two-color enhanced self-amplified spontaneous emission (SASE) [10] were proposed for attosecond pulse generation at FELs. A recently described scheme suggests possible generation of sub-attosecond pulses in the hard-X-ray region [11]. However, the stochastic pulse shape predicted for the pulses generated by these methods is disadvantageous. Furthermore, there are no reliable techniques available for CEP control of these or any other attosecond pulse sources. In contrast, recently we proposed a robust method for producing waveform-controlled linearly and circularly polarized CEP-stable attosecond pulses in the EUV spectral range [12, 13]. It uses relativistic electron bunches from a linear accelerator (LINAC), and relies on nanobunching by the inverse-FEL effect as well as on undulator radiation. In this setup waveform-controlled attosecond pulse generation is possible, unlike in other similar setups, where the predicted pulse energy and the short wavelength is imposing, but the waveform is stochastic [14].

In the present paper, a more detailed investigation is carried out on the feasibility of this technique. Practically important aspects were numerically studied. These include the dependence of the nanobunch length on the initial electron beam energy and energy spread, and on the modulator laser power. The dependence of the attosecond pulse energy on the radiation wavelength, initial electron energy, and undulator parameter were also studied. The shot-to-shot stability of the temporal shape (CEP) of the attosecond pulses is discussed, together with a possible way for isolated attosecond pulse generation.

## II. THE INVESTIGATED SETUP AND THE SIMULATION METHODS

The scheme of the setup proposed in Ref. [12] and further investigated in this work is shown in Fig. 1. A relativistic electron beam from a LINAC is sent through a modulator undulator (MU) where a TW-power laser beam is superimposed on it in order to generate nanobunches by the inverse free-electron laser (IFEL) action. Inside the MU the interaction between the electrons, the magnetic field of the undulator, and the electromagnetic field of the modulator laser introduces a periodic energy modulation of the electrons along the longitudinal ($z$) direction. This energy modulation leads to the formation of nanobunches in the drift space behind the MU. Typically, a drift space of only a few meters in length is required. Evidently, efficient generation of coherent and non-stochastic radiation pulses is possible only if the nanobunch length is shorter than the half period of the radiation. As our aim was to generate coherent EUV radiation in the 10 to 100 nm wavelength range, the minimization of the nanobunch length down to the sub-10-nm range was essential.

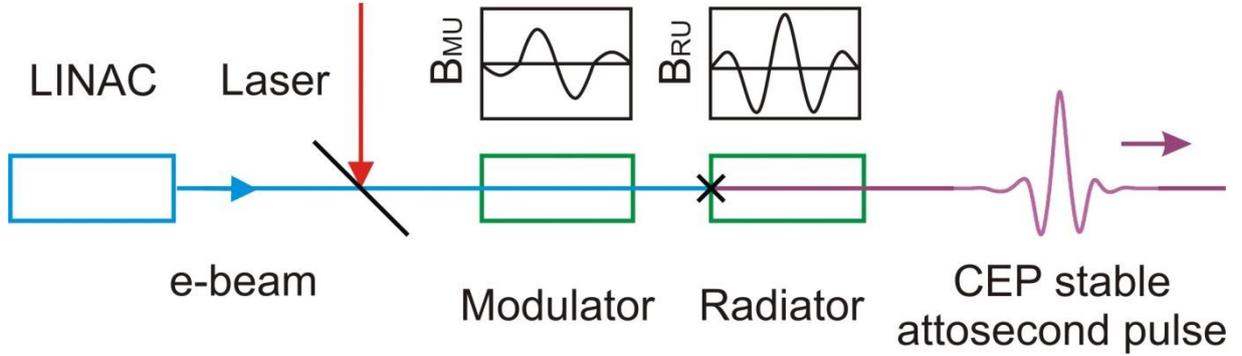

**Fig. 1.** Scheme of the investigated setup.

The nanobunched electron beam then passes through the radiator undulator (RU) consisting of a single period or of a few periods, depending on the desired waveform, and it emits electromagnetic radiation. The entrance of the RU is placed behind the MU at a position where the nanobunch length is the shortest. We refer to this position as the (temporal) focus ((x) symbol in Fig. 1). The wavelength of the generated radiation is determined by the well-known resonance condition [15]

$$\lambda_r = \frac{\lambda_{RU} \cdot \left(1 + K_{RU}^2/2\right)}{2\gamma^2}. \quad (1)$$

Here, $\lambda_r$ is the wavelength of the generated EUV radiation, $\lambda_{RU}$ is the period of the RU, $K_{RU} = eB_{0,RU}\lambda_{RU}/2\pi mc$ is the undulator parameter, $B_{0,RU}$ is the peak magnetic field of the RU, $e$ and $m$ are the electron charge and mass, respectively, $\gamma$ is the relativistic factor, and $c$ is the speed of light. The generated radiation waveform is mainly determined by the magnetic field distribution of the RU along the electron beam propagation direction $z$. For a better comparison of the XUV pulse parameters obtained in the different cases investigated below, the same RU magnetic field distribution

$$B_{RU} = \begin{cases} B_{0,RU} e^{-\frac{z^2}{2w_{RU}^2}} \cos\left(\frac{2\pi}{\lambda_{RU}}z\right), & \text{if } -\frac{L}{2} < z < \frac{L}{2}, \\ 0, & \text{otherwise} \end{cases} \quad (2)$$

was used throughout this work (see also Fig. 1). Here, $w_{RU}$ is the width of the Gaussian envelope and $L$ is the length of the RU. These parameters were set to $w_{RU} = 1.5 \times \lambda_{RU}$ and $L = 2.5 \times \lambda_{RU}$.

In order to consider a realistic situation, the initial electron bunch parameters (Table 1) were chosen according to published values for the accelerator of FLASH II at DESY in Germany [16,17]. In some cases, different parameter values were also used; these are specified below. We assumed a double-period MU, with antisymmetric magnet design and relative field amplitudes of -1/4, 3/4, -3/4, and 1/4. The modulator laser wavelength was $\lambda_L = 516$ nm. In this work, a similar theoretical model and simulation tools were used as in Ref. [12]. The General Particle Tracer (GPT) code was used for the numerical simulation of nanobunching in the MU [18]. A central longitudinal slice of the electron bunch with $3\lambda_L$ length was considered in the simulation, where the electrons were represented by macroparticles. 5000, in some cases 30000, macroparticles per bunch slice were used.

| Parameter | Value |
| --- | --- |
| E-beam energy ($\gamma$) | 2000 |
| E-beam intrinsic relative energy spread ($1\sigma_\gamma^*$) | 0.05 % |
| E-beam charge (total pulse) | $\approx 0.25$ nC |
| E-beam length | $\approx 30\ \mu m$ |
| E-beam normalized emittance | 1.4 mm mrad |
| E-beam radius | $30\ \mu m$ |
| Laser wavelength ($\lambda_L$) | 516 nm |
| Laser peak power | 10 TW |
| Laser beam waist inside MU | 0.72 mm |

Table 1. Parameters used in most of the simulations.

When the Coulomb interaction is neglected an approximate expression for the full width at half-maximum (FWHM) length of the nanobunch, $\Delta z_0$, can be given by analytical derivation [14,19]:

$$\Delta z_0 \approx 0.5 \frac{\lambda_L}{A}. \tag{3}$$

The relative energy modulation is given by $A = \Delta\gamma/\sigma_\gamma$, where $\sigma_\gamma = \sigma_\gamma^* \gamma$ is the intrinsic energy spread (Table 1) and $\Delta\gamma$ is the energy modulation acquired in the MU owing to IFEL action. At large relative modulation ($A \gg 1$) the nanobunch becomes much shorter than the laser wavelength.

The Coulomb interaction between the electrons limits the minimum achievable nanobunch length in the focus. This effect can be reduced by minimizing the drift length from the MU to the focus by using a chicane or higher laser power. In the simulations we used up to 10 TW modulator laser power, rather than a chichane, since the latter can cause distortions in the nanobunch. For the same reason, we also do not use any beam focusing elements, like quadrupole magnets.

TW-class table-top light sources with pulse durations comprising only a few optical cycles were intensely developed during the last few years [20,21]. For example, Herrmann et al. [20] reported a two-stage noncollinear optical parametric chirped-pulse amplification (OPCPA) system generating 16-TW, sub-three-cycle (130-mJ, 7.9-fs) pulses at an 805 nm central wavelength. Suitable light pulse sources are also being constructed elsewhere, for example in the Extreme Light Infrastructure (ELI) project [21]. Here we note that for $\lambda_r \gtrsim 20$ nm EUV wavelengths $\lambda_L \approx 800$ nm gives higher EUV pulse energies. Only below ~20 nm is the choice of a shorter laser wavelength (i.e. 516 nm, Table 1) advantageous, because of the shorter nanobunches. Though TW-class few-cycle OPCPA systems around 500 nm central wavelength need yet to be demonstrated, promising techniques are being developed [22,23,24]. In Section III.B we consider the possibilities for isolated attosecond EUV pulse generation, where laser pulses as short as two optical cycles are needed. For the generation of attosecond pulse trains longer laser pulses can be used.

We used the following handbook formula to calculate the electric field of the radiation generated in the RU [25]:

$$\vec{E}(t,\vec{r}) = \sum \left[ \frac{q\mu_0}{4\pi} \frac{\vec{R} \times (\vec{R} - R\vec{\beta}) \times \dot{\vec{v}}}{(R - \vec{R} \cdot \vec{\beta})^3} \right]_{ret}, \quad (4)$$

where $\mu_0$ is the vacuum permeability, $q$ is the macroparticle charge, $\vec{R}$ is the vector pointing from the position of the macroparticle at the retarded moment to the observation point, $\vec{v}$ is the velocity of the macroparticle, $\vec{\beta} = \vec{v}/c$, $c$ is the speed of light. The summation is for all macroparticles. During the radiation process the position, velocity, and acceleration of the macroparticles were traced numerically by taking into account the Lorentz force of the magnetic field of RU. The Coulomb interaction between the macroparticles was neglected during the undulator radiation process, because the transversal electron motion is by four orders of magnitude larger than the motion generated by Coulomb interaction.

## III. RESULTS

### A. Nanobunch generation

To find optimal conditions for nanobunching, the energy modulation in the MU was calculated as a function of both the undulator parameter ($K_{MU}$) and undulator period ($\lambda_{MU}$). The result is shown in Fig. 2 for $\gamma = 2000$ and $P_L = 10$ TW laser power. The figure also displays the $K_{MU}(\lambda_{MU})$ relation satisfying the resonance condition (similar to Eq. (1); black curve in Fig. 2). The highest obtained energy modulation was $\Delta\gamma = 145$, achieved at $K_{MU}=1.37$ and $\lambda_{MU} = 2.56$ m ((x) symbol in Fig. 2), being slightly off from the resonance condition. This offset is the effect of the Gouy phase, which cannot be neglected as the Rayleigh length of the Gaussian laser beam is comparable to the length of the MU. We note that the resonance condition assumes a plane-wave laser field. In our calculations $K_{MU}=1.4$ and $\lambda_{MU} = 2.08$ m were used ((+) symbol in Fig 2) as a possible trade-off between reducing the undulator length while still maintaining large energy modulation. In this case the energy modulation was $\Delta\gamma = 132$, which is about 9% smaller than the maximum $\Delta\gamma = 145$.

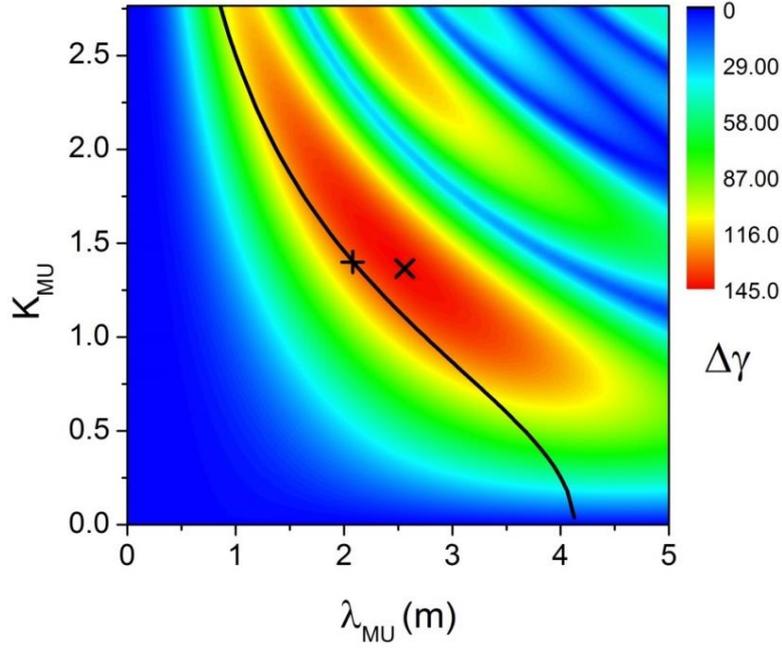

**Fig. 2.** Dependence of the energy modulation on the MU undulator parameter and undulator period for $\gamma = 2000$ and $P_L = 10$ TW laser power. The black curve indicates the resonance condition, the (x) symbol the maximum of the energy modulation, and the (+) symbol the parameters used in the calculations.

After determination of the energy modulation the nanobunch length at focus was investigated as function of the initial electron energy spread, as well as the electron energy and laser power. The dependence of the nanobunch length on the initial energy spread is shown in Fig 3. The relation is approximately linear, in accordance with Eq. 2. However, for small initial relative energy spreads below about $\sigma_\gamma^* = 0.15\%$ a significant deviation from the approximate linear dependence is observed. The reason is the increasing Coulomb interaction in this range, which does not allow to decrease the nanobunch length efficiently. Nevertheless, the lowest energy spread is the best for producing the shortest nanobunches.

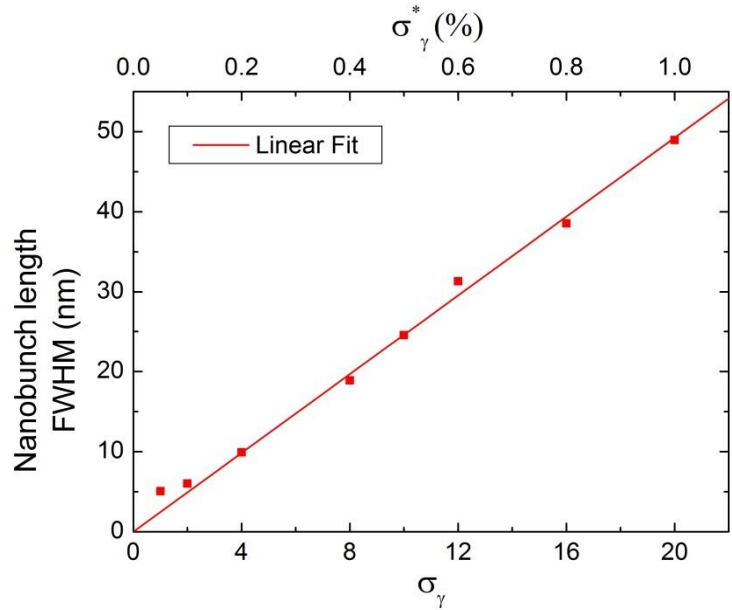

**Fig. 3.** Dependence of the FWHM nanobunch length, $\Delta z_0$, on the initial electron energy spread, $\sigma_\gamma$.

Typical relative energy spreads of electron bunches produced in LINACs are below 0.1%. According to Fig. 3, this should enable to produce nanobunches significantly shorter than 10 nm. This, in turn, should enable to generate radiation with as short as 10 nm wavelength and 40 asec duration (see Section 3.B). Laser-plasma based electron sources could be an alternative. However, at present their typical energy spread is in the range of 1% to 5% [26-28], which is far too high for EUV generation below 100 nm wavelength. Therefore we do not consider such sources here.

The nanobunch length was also investigated as function of the electron energy for a few different values of the modulator laser power (1, 2, 4 and 10 TW). The relativistic factor was varied between 1000 and 2000. The results of these calculations are shown in Fig 4. The figure contains the results of about 500 numerical simulation runs, determining also the error bars. The FWHM nanobunch length is displayed on a logarithmic scale for better visibility of the relative variation. The nanobunch length shows little dependence on $\gamma$, in accordance with Eq. (3). The shortest nanobunches were achieved with the highest laser power (10 TW). This is expected according to Eq. (3), as the highest laser field causes the highest energy modulation. Larger modulator laser power also decreases the length of the drift space, decreasing in this way the effect of the Coulomb repulsion. The charge of a single nanobunch is 1.1 pC and its length is as short as 6 nm at 4.9 m behind the center of the modulator undulator, with $\gamma=2000$ and 10 TW laser power. These nanobunches were used in the further calculations to obtain the EUV pulses described in Section 3.B. Nanobunches shorter than 10 nm can be generated by using higher than 4 TW modulator laser power (Fig. 4).

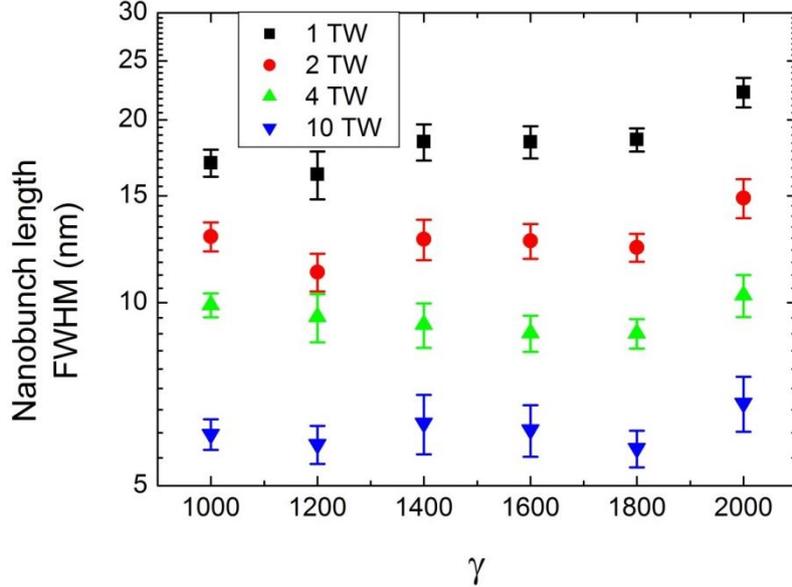

**Fig. 4.** Nanobunch lengths versus $\gamma$ for different values of the modulator laser power.

### B. EUV pulse generation

The temporal shape of the attosecond EUV pulses emitted by the extremely short electron nanobunches in the RU were calculated at a plane positioned 8 m behind the RU center. The emphasis was on exploring the dependence of the EUV pulse energy on important experimental parameters such as radiation wavelength, electron beam energy, and RU undulator parameter. As the EUV waveform can be conveniently set by the magnetic field distribution of the RU, similar scaling behavior can be expected for other waveforms.

The EUV pulse energy as function of the radiation wavelength ($\lambda_r$) in the range of 5÷250 nm is shown in Fig. 5a for $\gamma = 2000$, $P_L = 10$ TW, and two different RU undulator parameter values (0.5 and 0.8). The larger EUV pulse energy is obtained with the larger $K_{RU}$. The undulator parameter can be set to the desired value by adjusting the magnetic field amplitude. The radiation wavelength, given by the resonance condition Eq. (1), can be set by the choice of the RU period $\lambda_{RU}$. As seen in Fig. 5a, the pulse energy first increases with increasing wavelength, followed by saturation and subsequent energy decrease. The reason of the latter is the longer undulator period needed to generate longer wavelengths. Due to the associated longer path inside the RU the average nanobunch length and transversal size increases [13], thereby reducing coherence in the radiation process. As shown in Fig. 5b for 20 nm and 200 nm cases, the nanobunch lengths are increased by 33% and 1600% at the end of the radiator undulator (red and green curves), respectively. We note that for the generation of waveform-controlled pulses at longer wavelengths longer nanobunches (and lower modulator laser power, see Fig. 4) can be used more advantageously. In this case the relative change of the nanobunch length during propagation in the RU is reduced and a more favorable energy scaling can be achieved for long wavelengths.

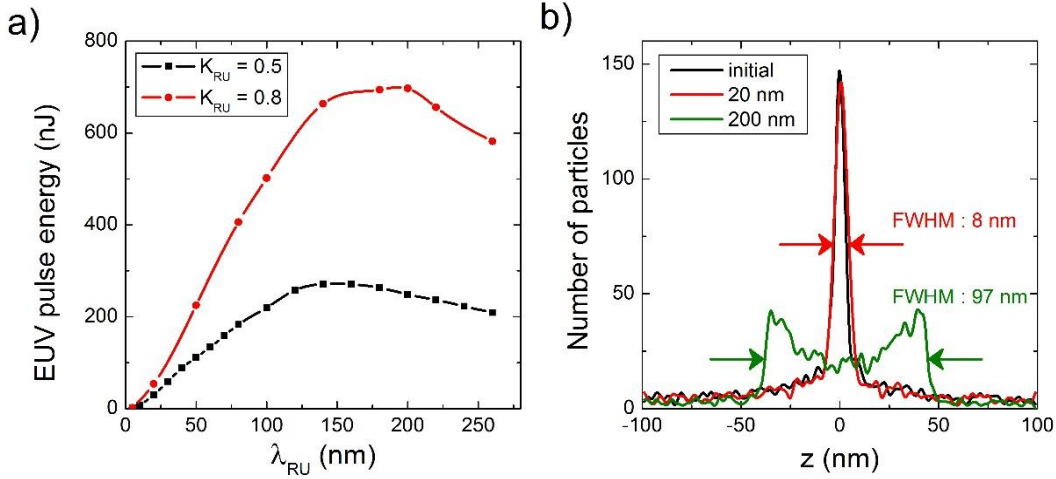

**Fig. 5.** (a) Dependence of the EUV pulse energy on the radiation wavelength $\lambda_r$ for $\gamma = 2000$ and $P_L = 10$ TW. (b) Distribution of macroparticles along the $z$ axis at the entrance (black curve) and the end of the RU for $\lambda_r = 20$ nm and 200 nm (red and green curves).

Increasing the energy of the attosecond EUV pulse is possible not only by increasing $K_{RU}$ but also by increasing $\gamma$. This is illustrated in Fig. 6 showing the dependence of the EUV pulse energy on $\gamma$ for $\lambda_r = 20$ nm and $K_{RU} = 0.5$. The energy at $\lambda_r = 20$ nm for $\gamma=2000$ and $K_{RU}=0.5$ is 23 nJ and the length of the RU and MU is 14.4 cm and 208 cm, respectively. There is a practical limit for increasing $\gamma$ further, because the period of the MU increases with $\gamma^2$, giving, for example, longer than 10 m period length for $\gamma=5000$.

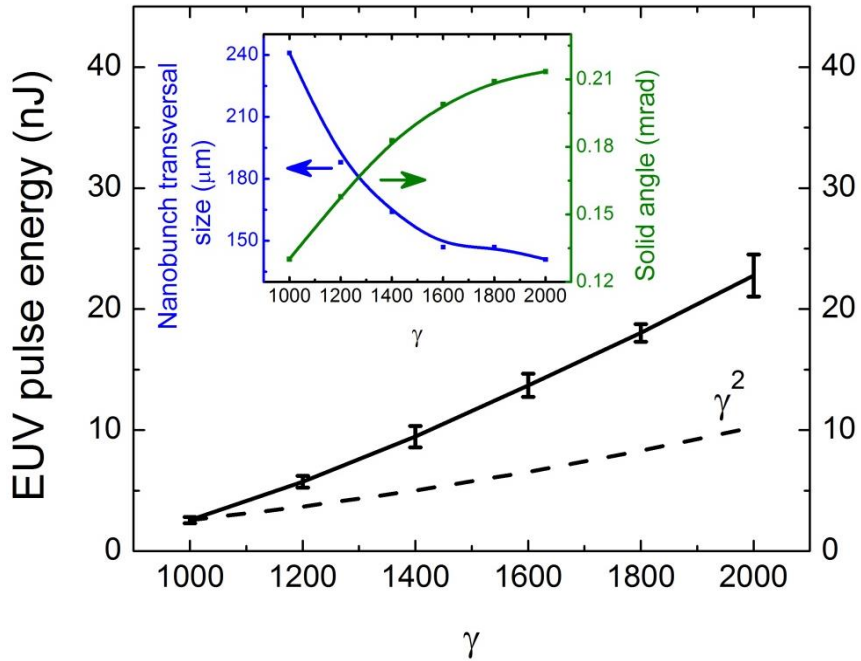

**Fig. 6.** Dependence of the EUV pulse energy on $\gamma$ for $\lambda_r = 20$ nm and $K_{RU} = 0.5$ (solid line). For comparison the dashed line indicates $\gamma^2$-dependence. The error bar represents the standard

deviation of the energy for different runs. The inset shows the nanobunch transversal size (blue curve) and the radiation solid angle (green curve).

Both the $K_{RU}$- and the $\gamma$-dependence of the EUV pulse energy can be explained by starting from the well-known formula [25]

$$N = \frac{2\pi}{3} \alpha K_{RU}^2 \qquad (5)$$

giving the number of photons, $N$, emitted by a single electron per undulator period ($\alpha$ is the fine-structure constant). This formula contains an explicit $K_{RU}^2$-dependence, while it does not depend on $\gamma$. However, for a nanobunch it implies a $\gamma$-dependence in the following way. The solid angle of the radiation emitted by a single relativistic electron is proportional to $1/\gamma^2$, and therefore the radiation fluence is proportional to $\gamma^2$. In case of a nanobunch the solid angle of radiation is determined by the transversal size of the nanobunch at the RU, rather than by the single-electron profile. The emission solid angle is reduced much below that of a single electron, as discussed in Ref. [12]. As the bunch transversal size decreases with increasing $\gamma$ (inset in Fig. 6, blue curve), the radiation solid angle increases with $\gamma$ (inset in Fig. 6, green curve). This, taken together with the $\gamma^2$-scaling of the radiation fluence of a single electron, results in EUV pulse energy increasing with $\gamma$ significantly faster than $\gamma^2$ (Fig. 6, solid and dashed lines).

In the third calculation series the RU undulator parameter was varied in the range of $K_{RU} = 0.1 \div 2.0$ and the $\lambda_{RU}$ undulator period was chosen such that for each value of $K_{RU}$ the radiation wavelength was kept at 20 nm and 60 nm, respectively. For both wavelengths, the EUV pulse energy is proportional to $K_{RU}^2$ (according to Eq. (4)) below about $K_{RU} = 0.6$, followed by saturation at larger $K_{RU}$ (Fig. 7(a)). The reason of the saturation is that a larger $K_{RU}$ results in more pronounced harmonics of the radiation wavelength and for shorter wavelength destructive interference occurs. Fig. 7(b) shows the spectra of the EUV pulses for three different $K_{RU}$ values for the 20-nm case. The fractional energy of the main spectral band decreases with increasing $K_{RU}$, as shown in the inset of Fig. 7(a). According to the calculations, single-cycle 80-as pulses with more than 60 nJ energy at 20 nm, and 240-as pulses with more than 450 nJ energy at 60 nm can be generated.

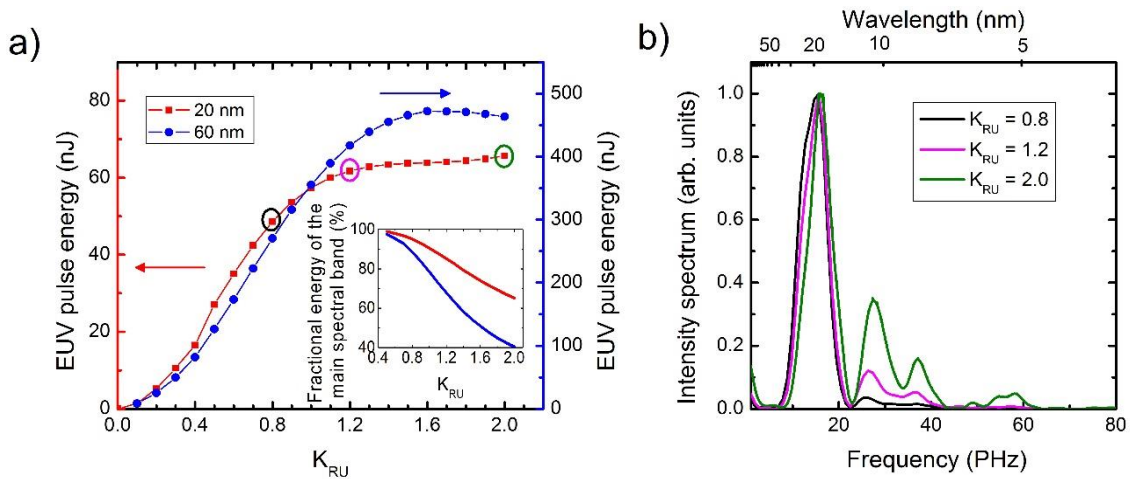

**Fig. 7.** (a) Dependence of the attosecond EUV pulse energy on RU undulator parameter at 20 nm (red) and 60 nm (blue) wavelengths. The inset shows the fractional energy of the main spectral band. (b) EUV pulse spectra for three different $K_{RU}$ values.

The EUV pulse energy as function of the modulator laser power was also investigated for different radiation wavelengths. For $K_{RU}=0.8$ the calculations predict the generation of attosecond pulses with 101 nJ, 177 nJ, and 270 nJ energy at 60 nm for 2 TW, 4 TW, and 10 TW modulator laser powers, respectively. We note that, the energy stability of the generated attosecond pulses depends on the stability of the modulator laser. In order to estimate this practically important aspect the power of the laser was varied by ±10%. According to our calculations with $\gamma = 2000$, $P_L = 10$ TW, and $\lambda_r = 20$ nm the energy fluctuation of the EUV pulse is 2.5 times higher than the fluctuation of the laser intensity.

Besides the pulse energy, it is the stability of the waveform which is of crucial importance for possible future attosecond field-driven experiments. Therefore, we also investigated the CEP-stability of the EUV pulses. The electron bunch length was kept constant for each simulation run, but the initial spatial distribution of the electrons was random. The temporal shapes of the generated attosecond pulses for each simulation run is displayed in the color-coded graphs of Fig. 8. The insets show the shapes of the EUV pulses along the respective dashed lines. As shown in this figure, the CEP fluctuation is very small. The standard deviations of the CEP are 31 and 13 mrad at 20 nm and 60 nm, respectively. We note that the CEP fluctuation of the attosecond pulses is by one order magnitude smaller in our setup than the CEP fluctuation of the most stable sources of few-cycle femtosecond pulses [29,30].

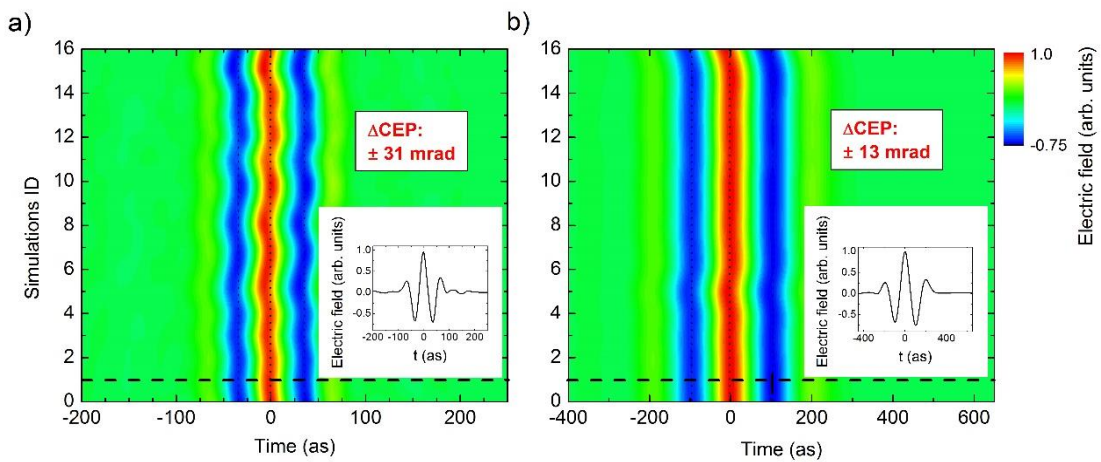

**Fig. 8.** EUV waveforms obtained for different random spatial electron distributions with otherwise identical parameters ($\gamma = 2000$, $P_L = 10$ TW, $K_{RU} = 0.5$) at 20 nm (a) and 60 nm (b) radiation wavelengths. The insets show the shapes of the EUV pulses corresponding to the respective dashed lines.

Since in our setup the electron bunch consists of multiple nanobunches separated by the modulation laser wavelength, a pulse sequence is generated. The ratio of separation time to pulse duration is smaller than 40. This can be too small for certain applications. One possibility for increasing this ratio substantially is to use two modulation lasers with significantly different wavelengths [10,31]. We investigated another possibility for isolated attosecond pulse generation, namely the shortening of the modulator pulse duration. In our simulations these pulses consisted of only a few optical cycles, and the phase of the carrier wave was adjusted such that the electric field was zero at the peak of the laser pulse envelope (sine pulse). Fig. 9 shows the calculated waveforms of the main attosecond pulse and the neighboring pulses for a few different modulator laser pulse durations. As the modulating pulse duration is decreased below about 5 cycles or 8 fs, the peak amplitude of the attosecond pulse preceding the central one starts to significantly decrease. Modulator pulses with less than about 2 cycles or 3.5 fs are able to generate isolated attosecond pulses.

Finally, we compare our results to the performance of attosecond EUV pulse sources based on high-order harmonic generation (HHG) in gas or plasma driven by high-intensity optical pulses [3,32]. Isolated attosecond pulses (IAPs) can be generated by few-cycle driving pulses or by using gating techniques with longer pulses [32]. Typical IAP energies are in the sub-nJ to 10 nJ range. The shortest pulse duration reported to date was 67 as; the wavelength range was ~10 to 20 nm [33]. CEP-stable single-cycle 130-as IAPs were reported with an estimated CEP fluctuation of 140 mrad [34]. However, no pulse energy was given in these works. Recently, the generation of 500-as IAPs with 1.3 µJ energy at ~30 eV was demonstrated [35]. Trains of attosecond pulses with an energy on the µJ scale can be generated by using many-cycle driving pulses. Significantly higher pulse and photon energies are expected from laser-produced plasmas [32,36], though the full potential of this method needs yet to be demonstrated.

In comparison, as shown above, similar shortest pulse durations can be achieved with our method as by HHG, at comparable wavelengths. However, unparalleled by any other source reported so far, our method easily enables the full control of the attosecond pulse waveform. The predicted CEP fluctuation is by one order of magnitude smaller than using HHG. For CEP-stable single-cycle attosecond pulses, the energy can be by one or two orders of magnitude higher in our case than that of HHG sources.

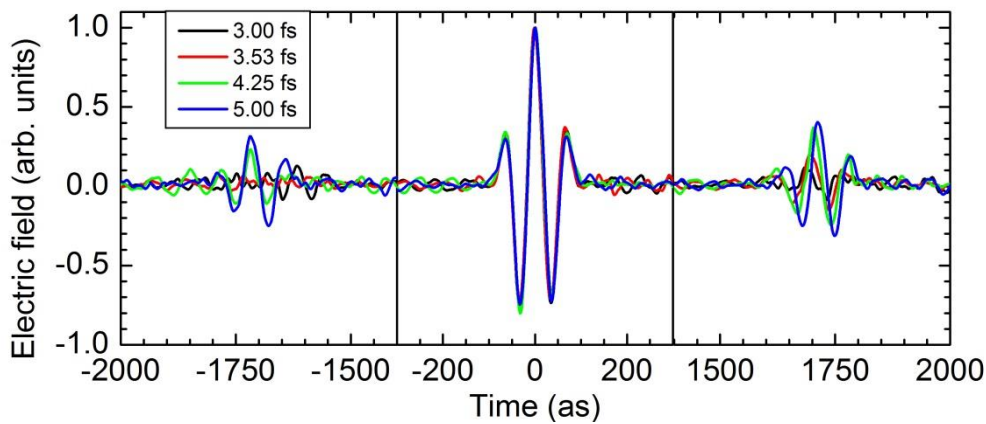

**Fig. 9.** Time dependence of the electric field of the generated attosecond pulse train for different modulator laser pulse lengths.

## IV. CONCLUSION

In summary, practical aspects of the method proposed in our previous work [12] for stable arbitrary-waveform attosecond EUV pulse generation were investigated in detail by means of numerical simulations. Optimal conditions for nanobunching were given and the dependence of the nanobunch length on initial electron beam parameters ($\gamma, \sigma_\gamma$) was investigated. Our calculations predict the generation of extremely short (<10 nm) electron nanobunches for high modulator laser powers (>4 TW).

The scaling of the generated attosecond EUV pulse energy with various parameters ($\lambda_r, \gamma, K_{RU}, P_L$) was studied. The CEP stability was discussed. For example, the nanobunches were predicted to emit 80-as EUV pulses at $\lambda_r$=20 nm with 23 nJ energy and ±31 mrad CEP stability, and 240-as pulses at $\lambda_r$=60 nm with 127 nJ energy and ±13 mrad CEP fluctuation for $K_{RU}$ =0.5. At longer generated wavelength or larger $K_{RU}$ the generation of EUV pulses with more than 500 nJ energy can be possible. The shortening of the modulator laser pulse duration was discussed for the generation of isolated attosecond pulses.

The proposed scheme can enable the development of practical sources of CEP stable attosecond EUV pulses using existing LINACs. These unprecedented pulses can be used for example in EUV pump—EUV probe experiments in the near future.

**ACKNOWLEDGEMENT**

Financial support from Hungarian Scientific Research Fund (OTKA) grant No. 113083 is acknowledged. JAF acknowledges support from János Bolyai Research Scholarship (Hungarian Academy of Sciences). The present scientific contribution is dedicated to the 650[th] anniversary of the foundation of University of Pécs, Hungary.